# Control of Photon Storage Time Using Phase Locking

**Byoung S. Ham**

*Center for Photon Information Processing, and the Graduate School of Information and Telecommunications*
*Inha University, 253 Yonghyun-dong, Nam-gu, Incheon 402-751, S. Korea*
*corresponding author: bham@inha.ac.kr*

**Abstract:** A photon echo storage-time extension protocol is presented by using a phase locking method in a three-level backward propagation scheme, where phase locking serves as a conditional stopper of the rephasing process in conventional two-pulse photon echoes. The backward propagation scheme solves the critical problems of extremely low retrieval efficiency and π rephasing pulse-caused spontaneous emission noise in photon echo based quantum memories. The physics of the storage time extension lies in the imminent population transfer from the excited state to an auxiliary spin state by a phase locking control pulse. We numerically demonstrate that the storage time is lengthened by spin dephasing time.

**1. Introduction**
Quantum interface is essential to quantum information processing, where flying qubits interact with an atomic medium [1-7]. The importance of quantum optical data storage varies, from applications of shorter storage time for quantum delay [8,9] in quantum computing to longer storage time of quantum memory [1-7] for quantum repeaters [10,11], which enable long-distance quantum communications. Recent observation of quantum interface using dynamic spectral gratings illustrates a modified version of the conventional photon echoes [12-14] to enhance retrieval efficiency in a dilute sample [1]. Unlike most other quantum memories limited by single mode storage [2-6], photon echoes use the reversible rephasing phenomenon in a collective atomic ensemble utilizing inhomogeneous broadening, so that consecutive multiple bit storage or multimode storage is possible [1,15-22]. Thus, from the practical standpoint, the quantum memory should be determined by: 1. Storage length of the quantum optical data; 2. Bandwidth, and 3. Retrieval efficiency. Regarding storage time, several modified photon echo techniques have resulted in storage time extension, replacing a constraint due to optical dephasing with spin dephasing, which is normally one order of magnitude longer. Examples include controlled reversible inhomogeneous broadening (CRIB) [15-18] and frequency comb-based spectral grating methods [15]. A recent proposal of resonant Raman echoes using optical locking shows a breakthrough in the storage time by extending it to spin population relaxation time [7]. Regarding echo efficiency, several modified methods have been presented so far: 1. A backward propagation scheme using an auxiliary spin state to overcome the geometry-based intrinsically low photon echo efficiency [16-18]; 2. Direct atom swapping by using a DC Stark field for rephasing [18,19]; and 3. Slow light-enhanced photon echoes to lengthen the light-matter interaction time for enhanced absorption, even in a dilute optical medium [21,22]. The slow light-enhanced photon echo observation in particular has demonstrated a few hundred times increased photon echo efficiency [22]. However, most of the modified versions of the photon echoes mentioned above require an additional process, which may be a drawback. Examples include: 1. External DC field usage for the rephasing process [18,19]; 2. A long optical pumping process for frequency comb generation [1,15]; 3. Ratio frequency usage for rephasing [17,20]; and 4. Requirement of electromagnetically induced transparency [23], which is difficult to obtain in most solid media [21].

In this Article we present a phase locked photon echo protocol to extend the photon storage time as well as to obtain near perfect retrieval efficiency. The storage time extension results from phase locking via an auxiliary spin state, where the phase locking temporally freezes the optical rephasing process by transferring the excited atoms to an auxiliary spin state. Thus, the storage time is now extended by the spin dephasing process of the auxiliary spin state. The distinctive difference of the present method from the CRIB is the optical control of the photon echo rephasing process without relying on external DC field and/or rf pulses. The backward propagation photon echo scheme has been originally introduced in conventional three-pulse photon echoes [13,14]; otherwise we meet a dilemma between data photon absorption and echo reabsorption [22,24]. Moreover, the backward propagation photon echoes have potential for aberration corrections when dealing with quantum imaging applications [25]. Regarding the controversial point of π rephasing pulse-induced spontaneous emission noise in conventional two-pulse photon echoes [26], the present technique lessens this controversy by using spectrally and spatially separable backward phase locked control pulse resulting in noncollinear echo propagation. Thus, the present phase locked photon echo protocol sustains all the benefits of photon echoes with the additional advantages of storage time extension up to the spin phase decay time for quantum memory applications, where the storage time is one order of magnitude longer and retrieval efficiency is near perfect due to the backward propagation scheme.

## 2. Theory and Discussions

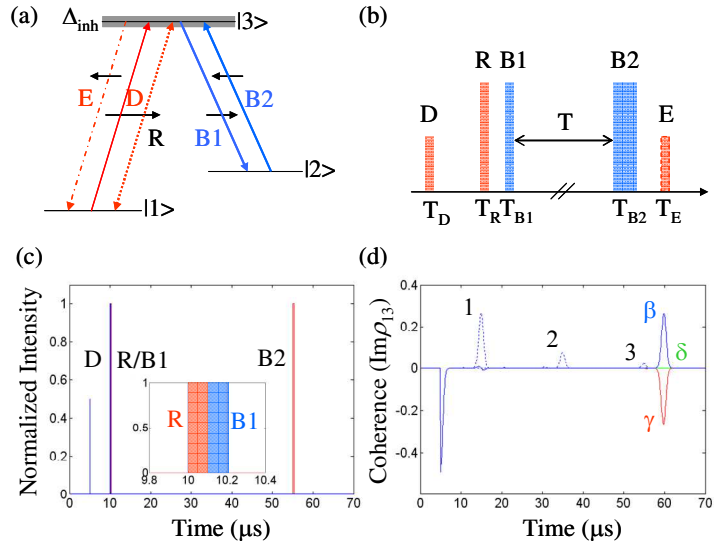

Fig. 1. Phase locked echo. (a) Energy level diagram interacting with light pulses. B1, B2, D, E, and R stand for locking, unlocking, data, photon echo, and rephasing pulses. The pulse areas of D, R, B1, and B2 are π/2, π, π, and 3π, respectively. $\Delta_{inh}$ (680 kHz at full width at half maximum) is an optical inhomogeneous width with Gaussian distribution. (b) Pulse sequence. (c) Programed pulse sequence. $T_D$=5; $T_R$=10; $T_{B1}$=10.1; $T_{B2}$=55 μs. (d) Numerical simulations for (c). Marks 1, 2, and 3 represent conventional photon echoes without B1 and B2 for $T_R$=10, 20, and 30 μs, respectively. Marks β, δ, and γ respectively represent phase locked photon echoes with B1 (π) and B2 for B2=3π, 2π, and π in pulse area. The pulse area is defined by $\int \Omega dt$. $\Gamma_{12}=\gamma_{12}=0$; $\Gamma_{13}=\Gamma_{23}=5$ kHz; $\gamma_{13}=\gamma_{23}=10$ kHz. $\Omega_D=\Omega_R=\Omega_{B1}=\Omega_{B2}=5$ MHz.

Figure 1 shows a theoretical demonstration of the present phase locked photon echo method for storage time extension with near perfect retrieval efficiency. In Fig. 1(a) the states |1> and |3> interacting with quantum field or weak classical light (D, data) satisfy the conventional photon echo scheme. The light B1 and B2 transient with an auxiliary spin state |2> are used for phase locking and unlocking of the photon echo process initiated by R (rephasing). Here the phase locking means a controlled stoppage of the rephasing process via a complete transfer of excited atoms on state |3> to state |2> using a π pulse area of B1. Figure 1(b)

represents the pulse sequence of the present phase locked photon echo protocol. The first two red bars, D and R, configure a conventional photon echo sequence. The pulse area of R should be π for complete rephasing. The blue bars, B1 and B2 respectively, represent phase locking and unlocking by temporally stopping and resuming the rephasing process initiated by R. As a result, photon echo E follows at:

$$T_E = T_{B2} + (T_R - T_D) - (T_{B1} - T_R). \quad (1)$$

Figure 1(c) represents a programmed pulse sequence in the numerical calculations, and results are shown in Fig. 1(d). To maximize the photon echo signal, D is set to be π/2. However, the echo efficiency is independent of the D pulse area [see Fig. 2(a)]. Quantum field treatment using very weak coherent light has already been discussed [27]. In Fig. 1c, however, we neglect the delay of B1 [$(T_{B1} - T_R) \ll T_R$] to directly compare the phase locked photon echoes with conventional photon echoes, so that $T_E - T_{B2}$ is the same as $T_R - T_D$ in Eq. (1), positioned at t=60 μs.

For the numerical calculations, nine time-dependent density matrix equations are numerically solved without any assumption under the rotating wave approximation. The density matrix approach is powerful in dealing with an ensemble system interacting with coherent laser fields, owing to statistical information as well as quantum mechanical information [28]. The equation of motion of the density matrix operator $\rho$ is determined from Schrödinger's equation [28]:

$$\frac{d\rho}{dt} = -\frac{i}{\hbar}[H, \rho] - \frac{1}{2}\{\Gamma, \rho\}, \quad (2)$$

where $\{\Gamma, \rho\}$ is $\Gamma\rho + \rho\Gamma$, $H$ is the Hamiltonian, $\hbar$ is the Planck's constant divided by $2\pi$, and $\Gamma$ is the decay rate. The density operator $\rho$ is defined by $\rho = |\Psi\rangle\langle\Psi|$, where $|\Psi\rangle$ is the state vector. By solving equation (2) the following are obtained as main coupled equations:

$$\frac{d\rho_{13}}{dt} = i\Omega_{13}(\rho_{33} - \rho_{11}) - i\Omega_{23}\rho_{12} - i(\delta_1 + \gamma_{13})\rho_{13}, \quad (3)$$

$$\frac{d\rho_{23}}{dt} = i\Omega_{23}(\rho_{33} - \rho_{22}) - i\Omega_{13}\rho_{21} - i(\delta_1 + \gamma_{23})\rho_{23}, \quad (4)$$

$$\frac{d\rho_{12}}{dt} = i\Omega_{13}\rho_{32} - i\Omega_{23}\rho_{13} - i(\delta_1 - \delta_2)\rho_{12} - \gamma_{12}\rho_{12}, \quad (5)$$

where $\Omega_{13}$ ($\Omega_{23}$) is D or R (B1 or B2) Rabi frequency, and $\gamma_{ij}$ is the phase decay rate between states |i⟩ and |j⟩ [see Fig. 1(a)]. For simplicity, spin transition is assumed to be homogeneous, and both optical transitions are inhomogeneously broadened by $\Delta_{inh}$=680 kHz (full width at half maximum, Gaussian distribution). The optical inhomogeneous broadening $\Delta_{inh}$ is equally divided by a 10 kHz space for 161 divisions in the calculations. The results are the same for a weak or strong field of D. For visual effect, we use maximum coherence excitation by D, which is π/2 pulse area. Rabi frequency of all pulses is set at 5 MHz.

Figure 1(d) shows numerically calculated results of the present phase locked photon echoes. The curves with marks "1," "2," and "3" represent conventional photon echoes without B1 and B2, where the echo efficiency exponentially drops according to the optical phase decay time $T_2^{Opt}$ [$T_2^{Opt} = 1/(2\pi\gamma_3)$=8 μs; $\gamma_3 = \gamma_{31} + \gamma_{32}$=20 kH]: exp($-t/T_2^{Opt}$). In contrast, the photon echo efficiency of β with B1 and B2 shows the same value as the conventional one marked by "1," if the pulse area of B2 satisfies 3π. The storage time extension by B2 is determined by spin dephasing between states |1⟩ and |2⟩, which is assumed zero in Fig. 1(d). If the pulse area of B2 does not exceed 2π, photon echo generation cannot be expected (see δ for 0% with 2π of B2 or γ for −100% with π of B2, as discussed in Fig. 5). For proper phase locking, B1 must be applied before the rephasing process completes [see Fig. 2(c)]. Although the optical dephasing is locked by B1, the transferred atoms should decay according to the spin dephasing [see Fig. 2(b) and Fig. 3]. The echo generation results from time-delayed nondegenerate four-wave mixing processes, where phase matching must be carefully considered. Thus, the propagation direction of the echo signal must be opposite to that of the data. Due to different frequencies between D and B1, the echo trajectory is not exactly the same as that of D: $\mathbf{k}_E=\mathbf{k}_D-\mathbf{k}_{B1}+\mathbf{k}_{B2}$, where $\mathbf{k}_i$ is the propagation vector of light i. Regarding the R induced spontaneous emission noise, the pencil-like beam trajectory in an optically dense medium minimizes random spontaneous emission out of the beam trajectory $\mathbf{k}_R$ ($//\mathbf{k}_D$). Thus, the controversy of spontaneous emission noise onto echo signal discussed in Ref. 26 may not be serious and can be solved technically.

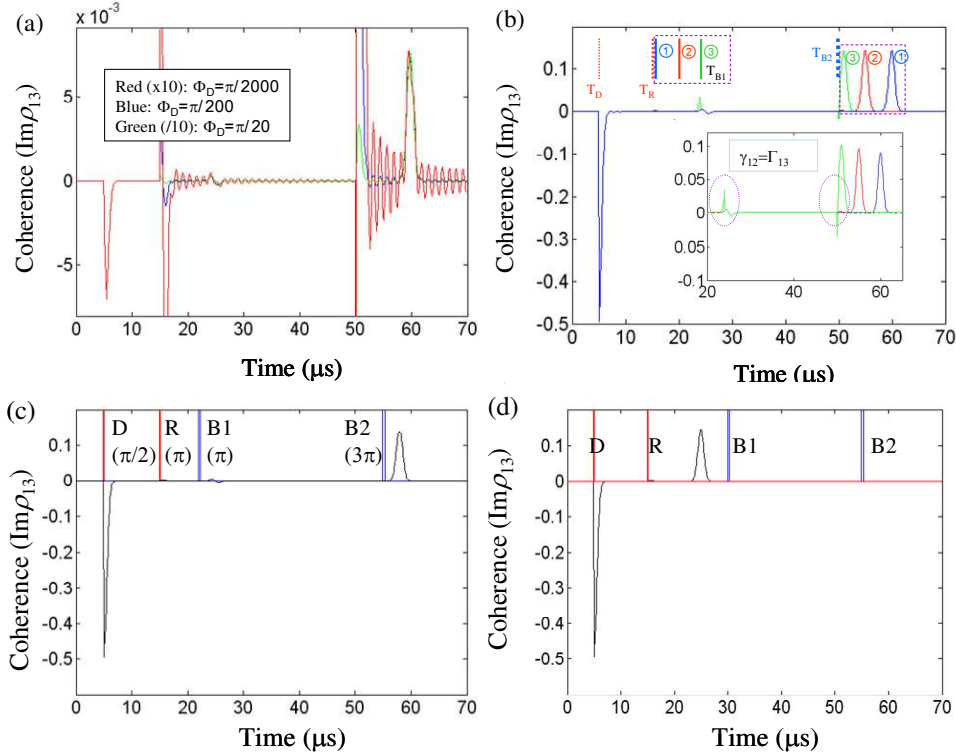

Fig. 2. (a) D-pulse area independent photon echo efficiency using phase locking. $\Phi_D$ stands for pulse area of the data pulse D. This linear relationship between efficiency and pulse area sustains up to $\Phi_D=\pi/10$. The data pulse D magnitude decreases by a factor of (i) 10 (green), (ii) 100 (blue), and (iii) 1000 (red), compared with that in Fig. 1. For comparison, the green (red) curve is multiplied (divided) by 10, while the blue curve shows no change. (b) B1 delay invariant phase locked echo. $T_{B1}$ in Fig. 1 and Eq. (1) is varied. The B1 delay from R is (c) 7 μs, and (d) 15 μs, while R delay from D is 10 μs. All other parameters are the same as in Fig. 1 except for $\gamma_{12}=\Gamma_{13}=2$ kHz in (a) and the inset of (b).

Figure 2(a) shows numerical calculations for the data pulse intensity invariant phase locked echoes. By decreasing the data field by a factor of 10, 100, and 1000 in the pulse area, the echo efficiency remains exactly the same. Figure 2(a), thus, implies a very important feature of quantum optical data storage in a collective atomic ensemble. In Fig. 2(b), Eq. (1) is numerically proved by controlling the delay time $T_{B1}$ of B1 from R. The delay $\Delta T$ of B1 from R for ①, ②, and ③ is $\Delta T=0$, 5, and 9 μs, respectively. The dotted circle in the inset indicates the case of ③ showing a small leakage of (rephased) echo due to very close timing of B1 to R. The gradual decrease of echo amplitude in the inset proves photon echo decoherence by spin phase decay rate $\gamma_{12}$. Figures 2(c) and 2(d) are without spin dephasing. With a 7 μs delaying of B1 from R, the echo appears at 58 μs, which is 3 μs delayed from B2: $T_E=T_{B2}+(T_R-T_D)-(T_{B1}-T_R)=55+(15-5)-(22-15)=58$. Figure 2(d) shows an extreme case of too much delay of B1 at t=30 μs. As shown, no phase locked echo appears, because the rephasing process completes before B1 switches on, by emitting a conventional two-pulse echo at t=25 μs. Thus, Fig. 2 concludes that, regardless of the delay ΔT of B1 from R, the phase locked echo keeps the same value as the conventional two-pulse photon echo with an additional storage time extension if spin dephasing is neglected (discussed in Fig. 3 in more detail).

Figure 3 shows phase locked echoes with 10 times increased spin dephasing $\gamma_{12}$ (spin phase decay only) than in the inset of Fig. 2(b). As shown, the phase locked echo severely decreases according to the spin dephasing as the storage time duration T between B1 and B2 increases. The decay feature of echo signals exactly matches the spin dephasing time: $\exp(-t/T_2^{Spin})$. Here the optical coherence excited by D is transferred into spin coherence via complete population transfer by B1. Thus, echo efficiency strongly depends on the spin dephasing process as shown in Fig. 3. In general the spin dephasing must include spin inhomogeneous broadening.

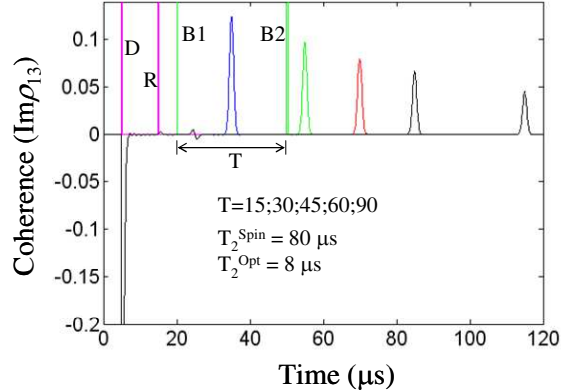

Fig. 3. Storage time versus spin phase decay. D(π/2), R(π), and B1(π) fall at t=5, 15, and 20 μs. B2 (3π) position at T=15 (Blue), 30 (Green), 45 (Red), 60 (Black), and 90 μs (last black peak). All other parameters are the same as in Fig. 1.

To achieve near 100% retrieval efficiency, however, a forward scheme of photon echo must be avoided due to reabsorption of the photon echoes by the residual atoms. This concern has been the main dilemma of photon echoes, because an optically dense medium is required for complete data absorption, while generated photon echoes experience more absorption as they propagate a longer distance due to the exponential decay of the data pulse absorption [24]. To solve this dilemma, a backward propagation scheme has been introduced in a modified photon echo scheme for nearly 100% echo efficiency [16-18]. Thus, complete photon echo recovery without reabsorption can be obtained. The backward propagation scheme in Fig. 1(a) acts as a quasi-phase conjugate, which is greatly beneficial when dealing with quantum images, where spatial aberration cancellation is expected [25,29,30].

Figure 4 shows analysis of the phase locking process presented in Figs. 1~3 in comparison with conventional two-pulse photon echoes. Figures 4(a) and 4(c) represent individual atom phase evolution. Each atom's phase velocity is determined by the detuning δ in the optical inhomogeneous broadening $\Delta_{inh}$. After the data pulse excitation, sum phases of all atoms decay out quickly, but the initial coherence is recovered by the rephasing pulse R. Figure 4(c) shows a symmetrically detuned atom pair (red and blue curves) to depict atom phase swapping by R at t=10 μs. This atom phase swapping is a fundamental characteristic of the time reversed process in photon echoes. Figures 4(b) and 4(d) show the evolution of phase locked atoms by B1 and B2 [for β in Fig. 1(d)], revealing exactly the same features as in Figs. 4(a) and 4(c), except for the storage time extension during the period T between B1 and B2.

This storage time extension is determined by the spin dephasing time of state |2>, which is set to zero for direct comparison.

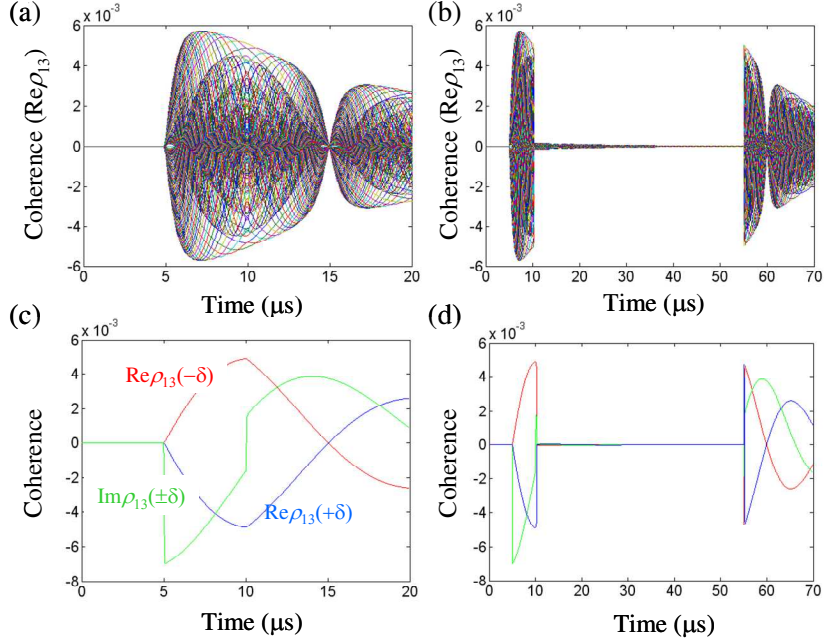

Fig. 4. (a) and (c) Conventional two-pulse photon echo. D (R) turns on at t=5 (10) μs. δ=40 kHz. (b) and (d) Phase locked photon echo using B1 (π) and B2 (3π). Here, 161 groups of atoms are considered for 1.6 MHz optical inhomogeneous broadening. All parameters are the same as in Fig. 1.

Figure 5 presents details of how the phase locking and unlocking play, as presented in Fig. 4(d). Figures 5(a) and 5(b) represent population density of each state, while Figs. 5(c) and 5(d) represent corresponding coherence. For the rephasing R (pink region), population swapping between states |1> and |3> by the π R pulse [see Fig. 5(a)] induces phase swapping between symmetrically detuned atoms by ±δ as discussed in Fig. 4(c). The π pulse of B1 induces population swapping between states |2> and |3> [see blue region of Fig. 5(a)], resulting in an additional π/2 phase shift between the symmetrically detuned atoms (±δ) [see blue region of Fig. 5(c)]. This π/2 phase shift ends up at zero coherence, resulting in freezing of phase evolution. This phase locking continues until the phase unlocking pulse B2 arrives. Because the population swapping between states |2> and |3> induces π/2 phase shift, it needs an additional 3π/2 phase shift to achieve the same phase recovery [see Figs. 4(c) and 4(d)], accomplished by the rephasing pulse R. This sequence is the key mechanism of the present method of phase locking. Unlike rephasing pulse R, the phase locking and unlocking pulses need twice the energy to shift the same amount of phase, because the pulse of B1 or B2 interacts with only state |3>, while R interacts with both states |1> and |3>.

Figures 5(e) and 5(f) depict Bloch vector (u, v) evolution in the uv plane for symmetrically detuned atoms by δ without [Fig. 5(e)] and with [Fig. 5(f)] phase locking and subsequent unlocking by B1 and B2. Dots with numbers denote timing of the pulse. As discussed in Fig. 4, the rephasing pulse R swaps symmetrically detuned atoms, so that the δ detuned atom (blue curve) after dot 2 must follow the −δ detuned atom trajectory (red curve) in a time reversed manner as shown in Fig. 5(e). In Fig. 5(f), the phase locking pulse B1 turned on at dot 3 (for +δ) renders the Bloch vector shrunk to zero upon reaching dot 4, meaning zero coherence magnitude in both absorption (Imρ$_{13}$) and dispersion (Reρ$_{13}$). Thus the phase evolution is locked until B2 arrives. The B2 pulse pumps atoms from state |2> back

to state |3>, resulting in absorption increase until complete depletion occurs by the π pulse ending at dot 6. Dot 6 is for γ in Fig. 1(d) leading to −100 % photon echo efficiency because it must follow the red curve in Fig. 5(f). Thus, to return to the same position as dot 3, an additional 2π pulse area is needed. This proves why B2 must be 3π. Therefore, we can conclude that the pulse area of R, B1, and B2 must satisfy the following equations for maximum photon echoes:

$$\Phi_R = (2n-1)\pi, \tag{6}$$
$$\Phi_{B2} = (4n-1)\pi \text{ for } \Phi_{B1} = (4n-3)\pi, \tag{7}$$
$$\Phi_{B2} = (4n-3)\pi \text{ for } \Phi_{B1} = (4n-1)\pi, \tag{8}$$
$$\Phi_{B1+B2} = 4n\pi, \tag{9}$$

where $\Phi_i$ is the pulse area of pulse *i*, and n is an integer (see Fig. 7).

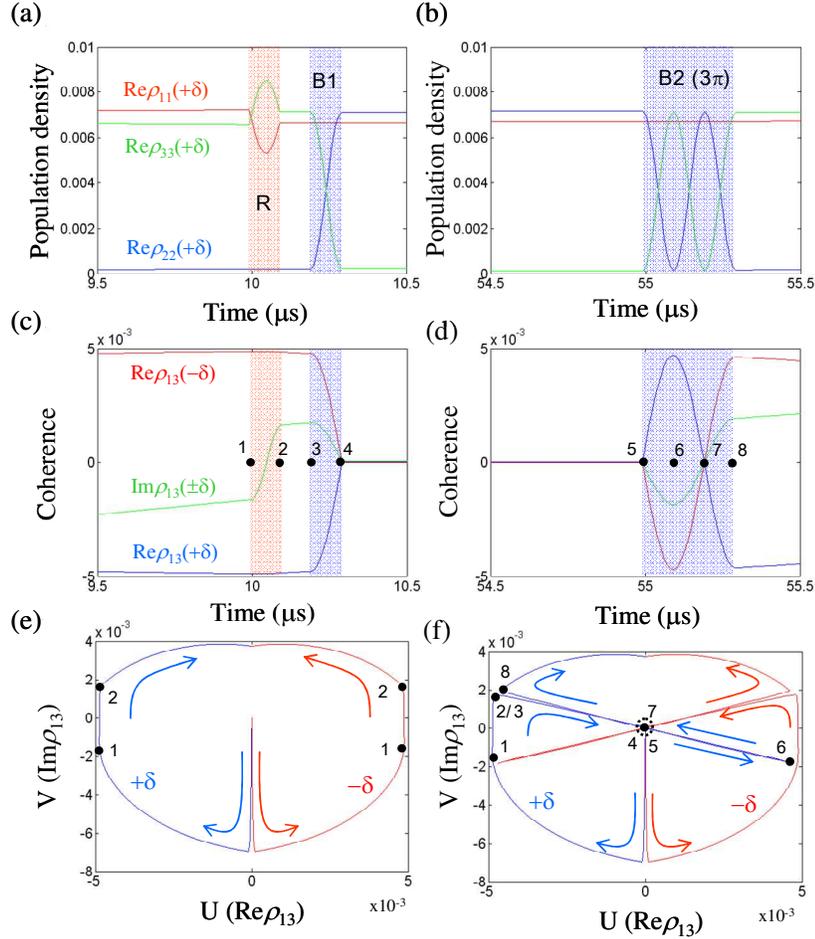

Fig. 5. Rephasing control by using B pulses. (a) and (b) Population evolution with R, B1, and B2 for symmetrically detuned atoms. (c) and (d) Coherence evolution with R, B1, and B2 for symmetrically detuned atoms. (e) and (f) Bloch vector evolution for (c) and (d), respectively. All parameters are the same as in Fig. 4, unless otherwise indicated.

Figure 6 shows details of phase swapping discussed in Fig. 5 with spin dephasing $\gamma_{21}$=2 kHz. Figures 6(c)~6(f) are to prove Eqs. (7) and (9). For a π pulse area of B1, 7π of B2 is also a solution. As shown in Fig. 7(d), the phase of Re$\rho_{13}$ is fully rephased when B2 satisfies 3π or

$7\pi$. Figures 6(e) and 6(f) represent Re$\rho_{13}$ and Im$\rho_{13}$ respectively for all atoms, where the phase of each atom is reversed every $2\pi$ (0.2 μs) pulse area of B2. Here, B2 turns on (off) at t=50.0 μs (50.7 μs).

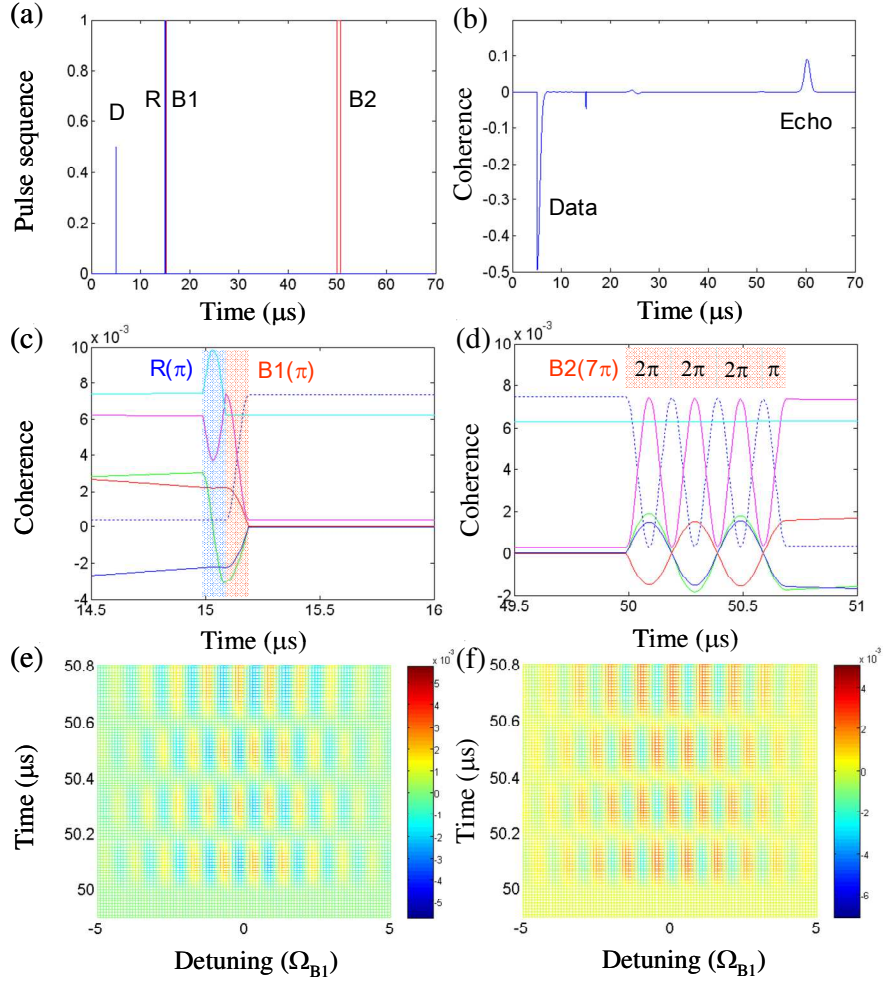

Fig. 6. Function of phase locking and unlocking pulses. (a) Pulse sequence of D($\pi$/2)-R($\pi$)-B1($\pi$)-B2($7\pi$). (b) Photon echo for (a). (c) and (d) Coherence evolution for symmetrically detuned atoms with R, B1, and B2. Red: Re$\rho_{13}(-\delta)$, Blue: Re$\rho_{13}(+\delta)$, Green: Im$\rho_{13}(\pm\delta)$, Cyan: $\rho_{11}$, Dotted: $\rho_{22}$, and Magenta: $\rho_{33}$. (e) Coherence evolution of Re$\rho_{13}$ versus detuning for B2. (f) Coherence evolution of Im$\rho_{13}$ versus detuning for B2. Parameters are the same as in Fig. 3. $\Omega_R=\Omega_{B1}=\Omega_{B2}$=5 MHz.

## 3. Conclusion

A phase locked photon echo protocol was proposed and numerically demonstrated, where conventional two-pulse photon echo storage time, limited by the optical phase decay process, is extended to spin dephasing time, which can be controlled using a magnetic field gradient. The backward propagation scheme in a three-level $\Lambda$–type system provides two important features for overcoming critical problems in conventional two-pulse photon echoes for quantum memory applications: near-perfect retrieval efficiency and spontaneous emission noise elimination due to a pencil-like slightly angled backward propagation geometry.

Extremely longer storage time with near perfect retrieval efficiency can also be obtained if the auxiliary spin state is isolated from the ground state.

**Acknowledgement**

This work was supported by the Creative Research Initiative Program (Center for Photon Information Processing) of the Korean Ministry of Education, Science, and Technology via National Research Foundation.